\documentstyle[11pt,moriond,epsfig]{article}

\def\Journal#1#2#3#4{{#1} {\bf #2}, #3 (#4)}

\def\PLB{{\em Phys. Lett.}  B}

\def\be{\begin{equation}}
\def\ee{\end{equation}}
\def\bea{\begin{eqnarray}}
\def\eea{\end{eqnarray}}

\begin{document}
\vspace*{4cm}
\title{LEP-WIDE COMBINATION OF THE SEARCH FOR THE STANDARD MODEL HIGGS BOSON}

\author{ A.N. OKPARA}

\address{Physikalisches Institut der Universit\"at Heidelberg, Philosophenweg 12,\\ 69120 Heidelberg, Germany}

\maketitle\abstracts{
A search for the standard model Higgs boson at LEP is reported. In the following, the results obtained by combining all standard model Higgs boson search channels of the experiments (at the end of run in the year 2000) are discussed. The data favour slightly a signal-and-background hypothesis. The excess over the background would be consistent with the existence of a standard model Higgs boson at $\approx 115~GeV$. A lower mass bound on the standard model Higgs boson is set to 113.5~GeV at a $95\%$~CL.}    

\section{Introduction}
A combined luminosity of approximately 830$~pb^{-1}$ was collected by the four experiments in the year 2000 at LEP. This note reviews the preliminary results obtained by combining all searches for the standard model Higgs boson, based on the status of the LEPC held in November 2000. The last week of data taking has therefore not been included and preliminary calibrations were used. The description of the individual analyses of the four experiments can be found in the following notes in these proceedings~\cite{pablo,enrico,gavin} and in later publications of the experiments~\cite{aleph,delphi,l3,opal}.

\section{Combination}

\begin{figure}[t]
\vspace*{-1.0cm}
\centerline{
\epsfig{file=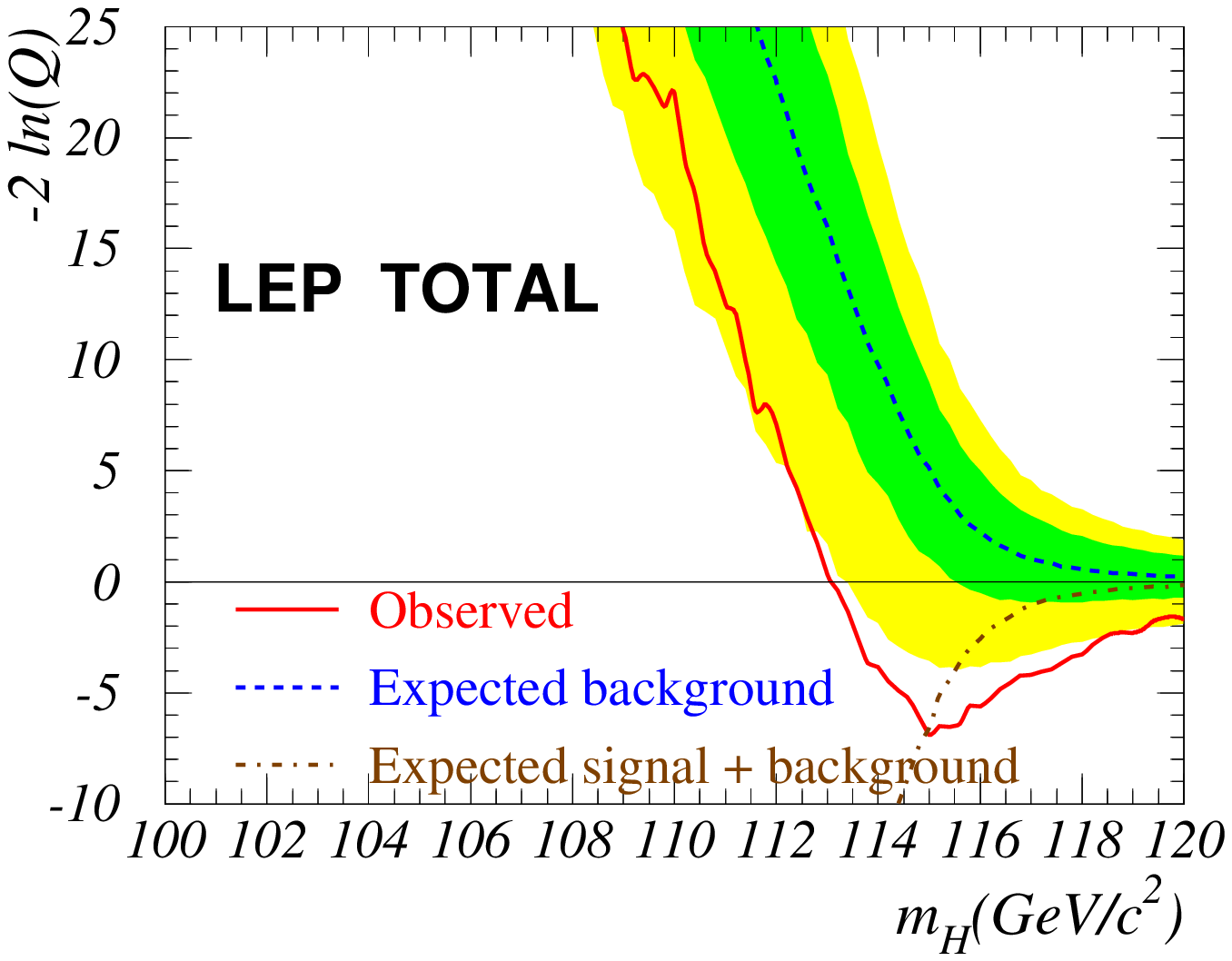,width=0.4\textwidth}
\hspace{0.15in}
\epsfig{file=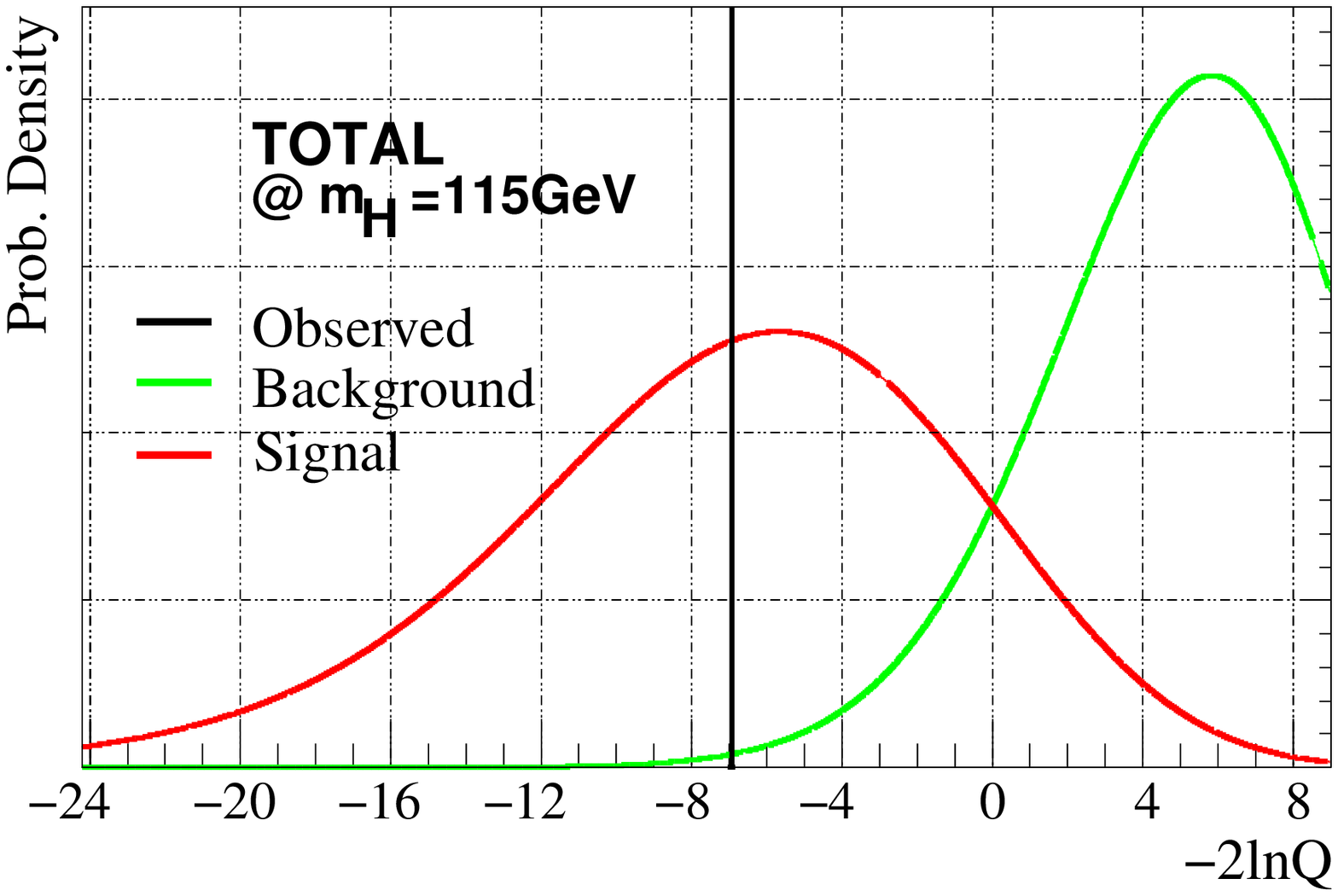,width=0.45\textwidth}
}
\caption{Left plot: Negative log-likelihood ratio as a function of $m_H$. Dashed line: median expected from the background-only hypothesis, with the one (dark grey) and two (light grey) standard deviation bands. Dashed-dotted line: expected from the signal-and-background hypothesis. Full line: observed in data. Right plot: Negative log-likelihood ratio for a $m_{H}=115~GeV$. Vertical line: observed in data. Light grey curve: expected from the background-only hypothesis. Dark grey curve: expected from the signal-and-background hypothesis. 
\label{fig:likeli}}
\end{figure}

\subsection{Statistical Method}

The data for different decay modes at different centre-of-mass energies have to be combined in order to obtain a statement on the probability of the existence for a possible signal. The information to the statistical analysis program is usually binned in a global variable, designed to discriminate signal against background events, which can contain information from the b-tag, kinematics or jet-properties. The expected background distribution $b_i$ (from MC computation), the expected signal $s_i$ (from MC computation) and the observed number of candidates $N_i$ is given as a function of bin number i. Using this information a likelihood ratio is used to create a {\it test-statistic (Q)}. It ranks the observed event configuration between the signal-and-background and the background-only hypothesis. Each bin $i$ is considered as a statistically independent measurement obeying Poisson statistics ($s_{tot}$ being the total expected signal rate). Because the signal spectra $s_i$, and possibly the background spectra $b_i$ and the observed number of candidates $N_i$, depend on the Higgs boson mass $m_H$ assumed for the analysis, the likelihood ratio is a function of $m_H$ too:

\begin{equation}
Q(m_H)={\mathcal{L}}(s+b)/{\mathcal{L}}(b) \qquad -2 lnQ(m_H)=2 s_{tot}-2\sum N_i ln[1+s_i/b_i]
\label{-2lnq}
\end{equation}

A confidence level $1-CL_b$ can be introduced as the probability that an arbitrary {\it Gedanken} experiment with background events gives a likelihood ratio $Q(m_H)$ higher than the observed one, it is therefore a measure for the compatibility of the data with the background-only hypothesis. $CL_{s+b}$ is likewise a measure of the compatibility of the data with the signal-and-background hypothesis. $CL_{s}=CL_{s+b}/CL_{b}$ can be used to set a lower mass bound of the standard model Higgs boson for a desired confidence level.

\section{Results}

\subsection{Log-likelihood ratio}

\begin{figure}[]
\centerline{
\psfig{figure=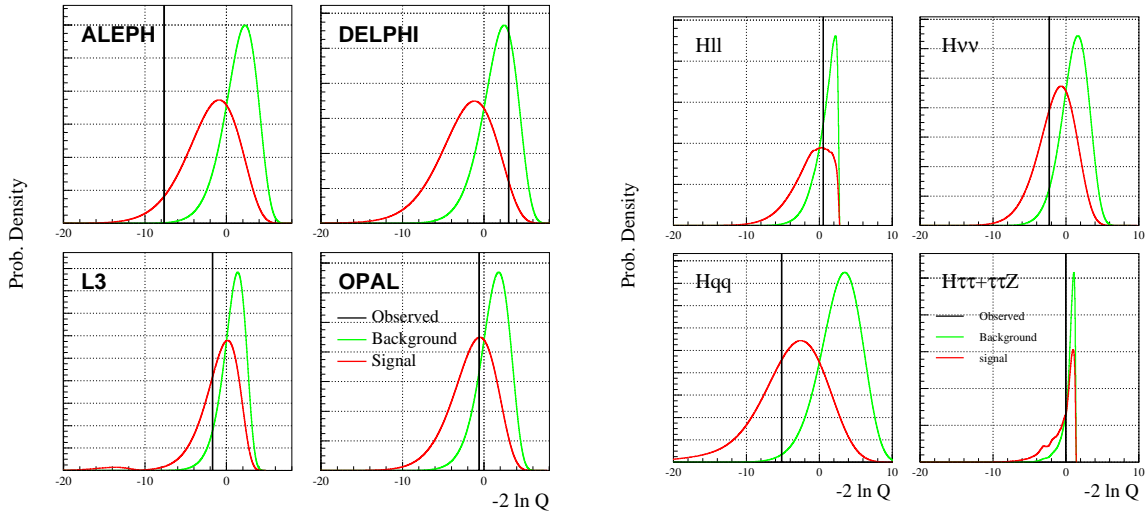,width=0.45\textwidth}
\hspace{0.15in}
\psfig{figure=wjm_4chan.eps,width=0.5\textwidth}
}
\caption{Left plot: The negative log-likelihood ratio for each experiment separately at $m_{H}=115~GeV$. Light grey lines: expected for the background-only hypotheses. Dark grey lines: expected in presence of a signal. Vertical lines: observed results. Right plot: The negative log-likelihood ratio for four channels at $m_{H}=115~GeV$.
\label{fig:likeliexp}}
\end{figure}

The left plot in figure~\ref{fig:likeli} shows the acquired negative log-likelihood ratio as a function of $m_H$ for the combined LEP results. The full curve is computed from observed data. The dashed curve is the expected negative log-likelihood ratio for the background-only hypothesis, it is shown together with the one (dark grey) and two standard deviation (light grey) bands. The dashed-dotted curve shows the expectation for the signal-and-background hypothesis. The maximal significance for a signal is derived from the minimum of the negative log-likelihood ratio. The position of the minimum is consistent with a standard model Higgs boson with mass of $\approx 115.0~GeV$. The right plot in figure~\ref{fig:likeli} shows a comparison of the log-likelihood ratio observed in the data (vertical line) with the distributions expected in the background-only (light grey curve) and the signal-and-background (dark grey curve) hypotheses, for a $m_{H}=115~GeV$ using the combined LEP data sample. The data clearly favour the signal-and-background hypothesis to the background-only hypothesis.

Figure~\ref{fig:likeliexp} shows results of the four experiments (left plot) and the four channels~\cite{pablo,gavin,enrico} (right plot) separately at $m_{H}=115~GeV$. The assignment of the curves is the same as in the combined plot. The comparison indicates that ALEPH favours the signal-and-background hypothesis over the background-only hypothesis. L3 and OPAL are both consistent with the signal-and-background hypothesis or the background-only hypothesis, while DELPHI seems to favour the background-only hypothesis. The combined four-jet channels (Hqq) and the missing energy channels (H$\nu\nu$) are more signal- than background-like while the lepton channels (H$\ell\ell$) and the tau channels (H$\tau\tau+\tau\tau $H) indicate no preferences.

\begin{table}[b]
\caption{Confidence level $1-CL_b$ at $m_H=115~GeV$ for the LEP experiments. Listed are the values obtained at the November LEPC.
}
\vspace{0.4cm}
\begin{center}
\begin{tabular}{|l|c||l|c||l|c||l|c|}
\hline
ALEPH   &   $1-CL_b$ &  DELPHI   &   $1-CL_b$ &  L3   &   $1-CL_b$ & OPAL   &   $1-CL_b$  \\
\hline
Total & 6.5$\times 10^{-4}$  & Total & 0.68 & Total & 6.8$\times 10^{-2}$ & Total & 1.9$\times 10^{-1}$\\
\hline
\end{tabular}
\label{tab:clbexp}
\end{center}
\end{table}

\begin{figure}[]
\begin{center}
\psfig{figure=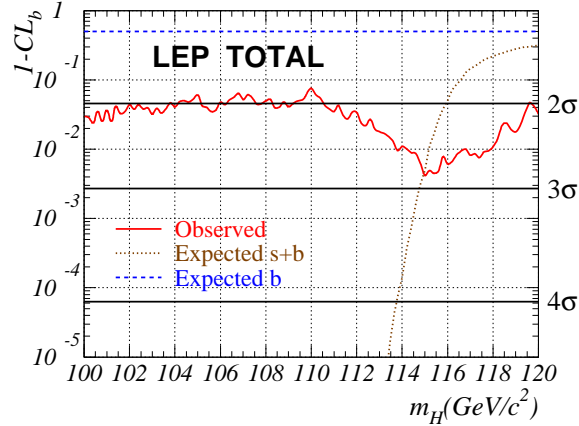,width=0.47\textwidth}
\end{center}
\caption{Confidence level $1-CL_b$ (full curve) as a function of $m_H$ for the combined LEP data sample. The horizontal lines indicate the levels for the $2 \sigma$, $3 \sigma$, $4 \sigma$ and $5 \sigma$  excesses above the expected background rate.\label{fig:clb}}
\end{figure}

\subsection{Confidence Level}

In table~\ref{tab:clbexp} the compatibility of the data with the background-only hypothesis ($1-CL_b$) at $m_H=115~GeV$ is presented for each experiment. Figure~\ref{fig:clb} shows the LEP combined compatibility as a function of the Higgs boson mass. The curve also shows a minimum at $115~GeV$ similar to the likelihood ratio curve. The probability that this arises from statistical fluctuation in the background is 0.42$\%$. With a confidence level of $95\%$ a lower bound on the standard model Higgs boson mass of $113.5~GeV$ is set, while a median value of $115.3~GeV$ is expected from background. 

\subsection{Contribution of single candidates}

A listing of the candidates and their weights ($s_i/b_i$) can be found in these proceedings~\cite{enrico,gavin}. The contributions of single candidates in bin $i$ can be easily assessed using equation~\ref{-2lnq}. It is simply $-2ln(1+s_i/b_i)$. A candidate having a high $s_i/b_i$ of 1.0 would therefore contribute $-$1.39 to the total test-statistic $-$2ln(Q). At a mass of $115~GeV$ the significance of the combined LEP search $1-CL_b$ results would decrease but still have the value of $\approx 0.9\%$ if such a candidate would not exist.

\section{Summary}

The standard model Higgs boson has been searched for at LEP. Using nearly all LEP2 data an excess over expected background has been observed which favours the signal hypothesis with the production of a standard model Higgs boson with a mass of $\approx 115~GeV$. A lower mass bound on the standard model Higgs boson of $113.5~GeV$ at $95\%$ is set, while $115.3~GeV$ would have been expected from background-only. The probability that the excess arises from statistical background fluctuation is $\approx0.42\%$. The significance is therefore to low to claim the discovery of the standard model Higgs boson.  A discovery could be claimed if the probability of such fluctuation would amount to less than $5.7$ x $10^{-7}$.

Meanwhile the four experiments published their results, including the last week of data taking and improved calibrations. Using these results a slight decrease in significance compared with the results presented in this note is expected. For a standard model Higgs boson mass of $115~GeV$ ALEPH~\cite{aleph} quotes a $1-CL_b$ of $2.7\times 10^{-3}$ and DELPHI~\cite{delphi} 0.77. L3~\cite{l3} has a slight increase to 9.0$\times 10^{-2}$ while OPAL~\cite{opal} remains nearly unchanged with 2.0$\times 10^{-1}$. A new LEP wide combination of the results will be done in future.

The aim of the LEP Higgs Working Group to run in the year 2001 and collect another $200~pb^{-1}$ of data per experiment at high energies ($\approx 208~GeV$ and above) was not accomplished. It has to be left to future experiments to ratify or confute the existence of the standard model Higgs boson at a mass of $\approx 115~GeV$.


\section*{References}
\begin{thebibliography}{99}

\bibitem{aleph}ALEPH Coll., \Journal{\PLB}{495}{1}{2000}.

\bibitem{delphi}DELPHI Coll., \Journal{\PLB}{499}{23}{2001}.

\bibitem{l3}L3 Coll., \Journal{\PLB}{495}{18}{2000}.

\bibitem{opal}OPAL Coll., \Journal{\PLB}{499}{38}{2001}.

\bibitem{pablo} P. Garcia, these proceedings. 

\bibitem{enrico} E. Piotto, these proceedings. 

\bibitem{gavin} G. Davies, these proceedings. 


\end{thebibliography}
\end{document}